\documentclass{kluwer}
\input{psfig.sty}

\begin{document}

\begin{article}
\begin{opening}
\title{Restart method and exponential acceleration of random 3-SAT 
instances resolutions: a large deviation analysis of the 
Davis--Putnam--Loveland--Logemann algorithm.}

\author{Simona \surname{Cocco} $^{1}$ and R{\'e}mi \surname{Monasson} $^{2,3}$}
\runningauthor{S. Cocco, R. Monasson}
\runningtitle{Restarts and Large deviations of DPLL}

\institute{$^1$  CNRS-Laboratoire de Dynamique des Fluides Complexes,
3 rue de l'Universit{\'e}, \\67000 Strasbourg, France.\\
$^{2}$ CNRS-Laboratoire de Physique Th{\'e}orique de l'ENS,
24 rue Lhomond, \\75005 Paris, France. \\
$^{3}$ CNRS-Laboratoire de Physique Th{\'e}orique,
3 rue de l'Universit\'e, \\67000 Strasbourg, France. }

\begin{abstract}
The analysis of the solving complexity of random 3-SAT instances using 
the Davis-Putnam-Loveland-Logemann (DPLL) algorithm slightly below threshold
is presented. While finding a solution for such instances demands
exponential effort with high probability, we show that an
exponentially small fraction of resolutions require a computation
scaling linearly in the size of the instance only. We compute
analytically this exponentially small probability of easy resolutions
from a large deviation analysis of DPLL with the Generalized
Unit Clause search heuristic,
and show that the corresponding exponent is smaller 
(in absolute value) than the growth exponent of the typical 
resolution time. Our study therefore gives some quantitative
basis to heuristic restart solving procedures, and suggests a natural
cut-off cost (the size of the instance) for the restart.
\end{abstract}
\keywords{restart, satisfiability, DPLL, large deviations.}
\end{opening}

\section{Introduction.}

Being a NP-complete problem, 3-SAT is not thought to be solvable in an
efficient way, {\em i.e.}  in time growing at most polynomially with
$N$. In practice, one therefore resorts to methods that need, {\em a
priori}, exponentially large computational resources. One of these
algorithms is the ubiquitous Davis--Putnam--Loveland--Logemann (DPLL)
solving procedure\cite{DP,survey}. DPLL is a complete search algorithm
based on backtracking.  The sequence of assignments of variables made
by DPLL in the course of instance solving can be represented as a
search tree, whose size $Q$ (number of nodes) is a convenient measure
of the instance hardness. Some examples of search trees are presented
in Figure~1.

\begin{figure}
\begin{center}
\psfig{figure=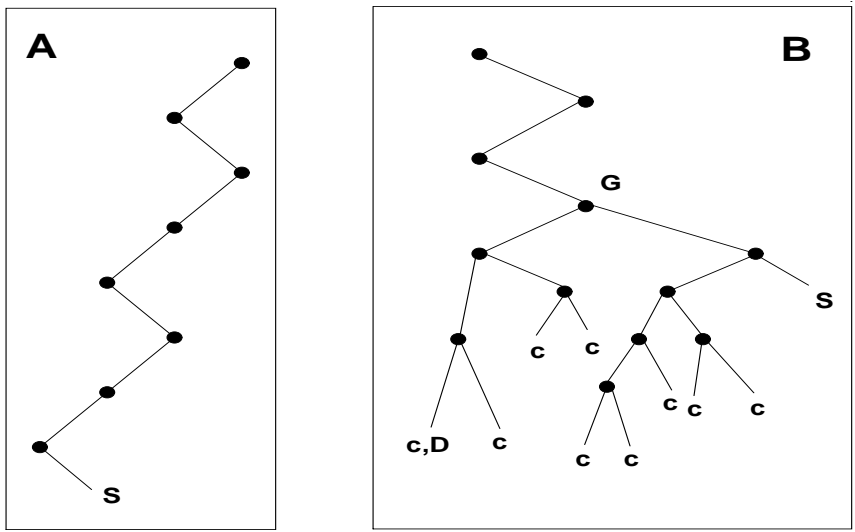,width=320pt,height=200pt,angle=-90}
\end{center}\vskip .5cm
\caption{Types of search trees generated by the DPLL solving procedure
on satisfiable instances. 
{\bf A.} {\em lower sat phase, $\alpha < \alpha _L$:} the algorithm finds
easily a solution with almost no backtracking. {\bf B.} {\em
upper sat phase, $\alpha _L < \alpha < \alpha _C$:} 
many contradictions (c) arise before
reaching a solution, and backtracking enters massively into play. 
Junction G is the highest node in the tree reached back by DPLL.
D denotes the first contradiction detected by DPLL, located at the
leaf of the first descent in the tree.}
\label{tree}
\end{figure}

In the past few years, many experimental and theoretical progresses
have been made on the probabilistic analysis of 3-SAT\cite{AI,Hans}.
Distributions of random instances controlled by few parameters are
particularly useful in shedding light on the onset of complexity. An
example that has attracted a lot of attention is
random 3-SAT: the three literals in a clause are randomly chosen  
variables, or their negations with equal probabilities, among a set of 
$N$ Boolean variables; clauses are drawn independently of each other.
Experiments\cite{AI,Mit,Cra,Kir} and theory\cite{Friedgut,Dubtcs}
indicate that clauses can almost surely always (respectively never) be
simultaneously satisfied if $\alpha$ is smaller (resp. larger) than a
critical threshold $\alpha _C \simeq 4.3$ as soon as $M,N$ go to
infinity at fixed ratio $\alpha$. This phase transition\cite{Sta} is
accompanied by a drastic peak in hardness at
threshold\cite{AI,Mit,Cra}. The emerging pattern of complexity is as
follows. At small ratios $\alpha < \alpha _L$, where $\alpha _L$
depends on the heuristic used by DPLL, instances are a.s. satisfiable
(sat). Finding a solution requires a tree whose size $Q$ scales only
linearly with the size $N$, and almost no backtracking is present
(Figure \ref{tree}A). Above the critical ratio, instances are
a.s. unsatisfiable (unsat) and proofs of refutation are obtained
through massive backtracking, leading to an exponential hardness $Q =
2^{N\omega}$ with $\omega >0$\cite{Chv,Bea}.

Recently, a quantitative understanding of the pattern of complexity
was proposed to estimate $\omega$ in the unsat regime as a function of
the ratio $\alpha$ of clauses per variables of the 3-SAT instance to 
be solved, and to unveil the structure of DPLL's search tree in the 
upper sat phase (Figure~\ref{tree}B), {\em i.e.} for ratios 
$\alpha _L < \alpha < \alpha _C$ \cite{Coc1}.  
In the latter range, instances are a.s. sat, but
their resolution requires with high probability an exponentially large
computational effort $(\omega >0)$ \cite{Coc1,Vardi1,Vardi2,Achl3}.
Theoretical predictions for $\omega$ as a function of $\alpha$ were
derived \cite{Coc1}, extending to the upper sat phase the calculations
of the unsat phase.

In this paper, we study in more detail the upper sat phase, and more
precisely the distribution of resolutions complexities of randomly
drawn instances with ratios $\alpha _L < \alpha < \alpha _C$. Using
numerical experiments and analytical calculations, we show that,
though complexity $Q$ a.s. grows as $2^{N\omega}$,
there is a finite but exponentially small probability $2^{-N
\zeta}$ that $Q$ is bounded from above by $N$ only.  In other words,
while finding solutions to these sat instances is almost always
exponentially hard, it is very rarely easy (polynomial time). Taking
advantage of the fact that $\zeta$ is smaller than $\omega$, we show
how systematic restarts of the heuristic may decrease substantially
the overall search cost. Our study therefore gives some theoretical
basis to incomplete restart techniques known to be efficient to solve
satisfiable instances\cite{Dubois,Gomes}, and suggests a natural
cut-off cost for the restart.

We first start by recalling our previous framework for studying resolutions
taking place with high probability, which is helpful to understand rare 
resolutions too (Section II). Numerical experiments are presented in 
Section III. Section IV is devoted to the analytical calculation of 
$\zeta$, and we present some conclusions in Section V. We have tried to
highlight the different status of the statements
and results presented: rigorous, expected to be exact
but proof lacking, approximate, or experimental. We hope
this effort will benefit the reader and make our work more accessible.

\section{Resolution trajectories: the high probability scenario.}

In this section, we briefly recall the main features of the resolution
by DPLL of satisfiable instances of size $N$, occurring with large
probability as $N\to\infty$\cite{fra2,Fra,Achl,Coc1}. 

\begin{center}
\begin{figure}
\psfig{figure=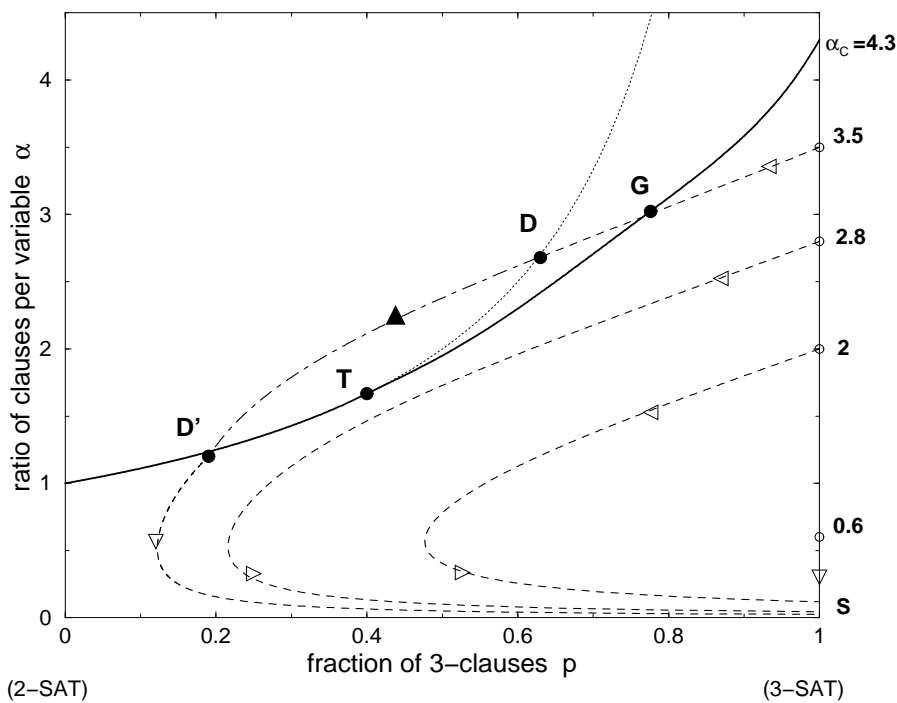,width=340pt,angle=-90}
\vskip .5cm
\caption{Phase diagram of 2+p-SAT and dynamical trajectories
of DPLL for satisfiable instances.
The threshold line $\alpha_C (p)$ (bold full line) separates sat
(lower part of the plane) from unsat (upper part) phases. Extremities
lie on the vertical 2-SAT (left) and 3-SAT (right) axis at coordinates
($p=0,\alpha _C=1$) and ($p=1,\alpha _C\simeq 4.3$) respectively.  
Departure points for DPLL trajectories are located on the
3-SAT vertical axis  (empty circles) 
and the corresponding values of $\alpha$ are explicitly given. 
Arrows indicate the direction of "motion" along trajectories (dashed curves)
parametrized by the fraction $t$ of variables set by DPLL.  For small
ratios $\alpha < \alpha _L$, branch trajectories remain
confined in the sat phase, end in S of coordinates $(1,0)$, where a
solution is found. At $\alpha_L$ ($\simeq 3.003$ for the GUC heuristic, 
see text), the single branch trajectory hits
tangentially the threshold line in T of coordinates $(2/5,5/3)$. In
the range $\alpha _L < \alpha < \alpha_C$, the branch
trajectory intersects the threshold line at some point $G$ (that depends
on $\alpha$). With high probability, a contradiction arises before the 
trajectory crosses the dotted curve $\alpha =1/(1-p)$ (point D);
through extensive backtracking, DPLL later reachs back the highest 
backtracking node in the search tree ($G$) and find a solution
at the end of a new descending branch, see Figure~1B. With
exponentially small probability, the trajectory  (dot-dashed curve, full
arrow) is able to cross 
the "dangerous" region where contradictions are likely to occur; it
then exits from this contradictory region (point $D'$) and
ends up with a solution (lowest dashed curve, light arrow).}
\label{diag}
\end{figure}
\end{center}

The action of DPLL on an instance of 3-SAT causes the changes of the
overall numbers of variables and clauses, and thus of the ratio
$\alpha$.  Furthermore, DPLL reduces some 3-clauses to 2-clauses. We
use a mixed 2+p-SAT distribution\cite{Sta}, where $p$ is the fraction
of 3-clauses, to model what remains of the input instance at a node of
the search tree. Using experiments and methods from statistical
mechanics\cite{Sta}, the threshold line $\alpha _C (p)$, separating
sat from unsat phases, may be estimated with the results shown in
Figure~\ref{diag}. For $p \le p_T = 2/5$, {\em i.e.} left to point $T$, 
the threshold line is given by $\alpha _C(p)=1/(1-p)$, and saturates 
the upper bound for the satisfaction of 2-clauses\cite{Sta,Achl}. 
Above $p_T$, no exact expression for $\alpha _C (p)$ is known.

The phase diagram of 2+p-SAT is the natural space in which the DPLL
dynamic takes place. An input 3-SAT instance with ratio $\alpha$ shows
up on the right vertical boundary of Figure~\ref{diag} as a point of
coordinates $(p=1,\alpha )$.  Under the action of DPLL, the
representative point moves aside from the 3-SAT axis and follows a
trajectory which depends on the splitting heuristic implemented in
DPLL. We consider here the so-called Generalized Unit-Clause (GUC)
heuristic proposed by Franco and Chao\shortcite{fra2,Fra}
\cite{Francotcs,Achl}.  Literals
are picked up randomly among one of the shortest available
clauses. This heuristic does not induce any bias nor correlation in
the instances distribution\cite{fra2}. Such a statistical
``invariance'' is required to ensure that the dynamical evolution
generated by DPLL remains confined to the phase diagram of
Figure~\ref{diag}.

Chao and Franco were able to analyze rigorously  resolutions
corresponding to initial ratios $\alpha < \alpha _L \simeq
3.003$. Their analysis consists in monitoring the evolution of the
densities (numbers divided by $N$) $c_2$ and $c_3$ of 2- and 3-clauses 
respectively as more and more variables are assigned by DPLL. Both
densities become highly concentrated around the averages as the size
$N$ goes to infinity. Calling $t$ the fraction of assigned variables, 
$c_2(t)$ and $c_3(t)$ obeys a set of coupled ordinary differential
equations (ODE),
\begin{eqnarray}
\frac {d c_3 (t)} {dt} &=& - \frac{3}{1-t} c_3 (t) - \rho _3 (t)
\nonumber \\
\frac {d c_2 (t)} {dt} &=& \frac{3}{2(1-t)} c_3 (t) - 
\frac{2}{1-t} c_2 (t) - \rho _2 (t) \quad ,
\label{diff}
\end{eqnarray}
where $\rho_2 (t), \rho_3(t)$ are the probabilities that the split is
made from a 2-, 3-clause respectively. For GUC and an initial ratio
$\alpha_0 >2/3$, $\rho_2(t) = 1 - c_2(t)/(1-t), \rho_3(t)=0$. 

To obtain the single branch trajectory in the phase diagram of Figure~2,
we solve the ODEs (\ref{diff}) with initial conditions $c_2(0)=0, 
c_3(0)=\alpha_0$, and perform the change of variables
\begin{equation}
\alpha (t) = \frac{c_2(t)+c_3(t)}{1-t} \qquad , \qquad
p(t) = \frac{c_3(t)}{c_2(t)+c_3(t)} \qquad , 
 \label{change}
\end{equation}
to obtain
\begin{equation}
\alpha (t) =\frac{\alpha _0}4 (1-t)^2 + \frac {3\alpha _0}4
+\ln (1-t) \quad , \quad
p (t) =  \frac{\alpha _0 (1-t)^2 }{\alpha (t)}
 \quad .
\label{diff6}
\end{equation}

Results are shown for the GUC heuristics and starting ratios $\alpha_0 =2$
and 2.8 in Figure~\ref{diag}. The trajectory,
indicated with a dashed line, first heads to the left and then
reverses to the right until reaching a point on the 3-SAT axis at
a small ratio. Further action of
DPLL leads to a rapid elimination of the remaining clauses and the
trajectory ends up at the right lower corner S, where a solution is
found. Note that for initial ratios $\alpha _0<2/3$, only the second
part of the trajectory restricted to the $p=1$ axis subsists.

Frieze and Suen \shortcite{Fri} 
have shown that, for ratios $\alpha _0 < \alpha _L \simeq 3.003$
(for the GUC heuristics), the full search tree essentially reduces
to a single branch, and is thus entirely described by the ODEs (\ref{diff}).
The amount of backtracking necessary to reach a solution is bounded from above 
by a power of $\log N$. The average size of the branch $Q$ scales linearly
with $N$ with a multiplicative factor $\gamma (\alpha _0)=Q/N$ that can
be computed exactly\cite{Coc1}.

The boundary $\alpha _L$ of this easy sat region can be defined as the largest
initial ratio $\alpha _0$ such that the branch trajectory $p(t),\alpha (t)$ 
issued from $\alpha _0$ never leaves the sat phase in the course of DPLL 
resolution. Indeed, as $\alpha _0$ increases up to $\alpha_L$, 
the trajectory gets closer and closer to the threshold line $\alpha _C 
(p)$. Finally, at $\alpha _L \simeq 3.003$, the trajectory touches the 
threshold curve tangentially at point $T$.

The concept of trajectory helps to understand how resolution takes
place in the upper sat phase, that is for ratios $\alpha _0$ ranging 
from $\alpha _L$ to $\alpha _C$. The branch trajectory,
started from the point $(p=1,\alpha _0)$ corresponding to the initial
3-SAT instance, hits the critical line $\alpha_c(p)$ 
at some point G with coordinates ($p_G,\alpha_G$) after $N\;t_G$ variables 
have been assigned by DPLL, see Figure~\ref{diag}.  The
algorithm then enters the unsat phase and generates 2+p-SAT instances
with no solution. Backtracking will appear as soon as a contradiction
is detected by DPLL. This may occur at any point along the trajectory
\cite{Fri}, but no further than the crossing point $D$ with
the $\alpha=1/(1-p)$ line (beyond $D$, unit-clauses are created at a
rate larger than their elimination through unit-propagation, and 
opposite literals will appear w.h.p.). Later, 
massive backtracking enters into play
\cite{Coc1} until $G$ is reached back by DPLL. $G$ is indeed the
highest backtracking node in the tree, since nodes above $G$ 
are located in the sat phase and carry 2+p-SAT instances with 
solutions (Figure~\ref{tree}B). DPLL will eventually reach
a solution $S$ (Figure~\ref{tree}B).

\section{Numerical experiments.}

In this section we present some numerical experiments on large but finite
instance sizes, showing some deviations from the high probability 
scenario exposed above.

\subsection{Instance-to-instance distribution of complexities.}

\begin{center}
\begin{figure}
\psfig{figure=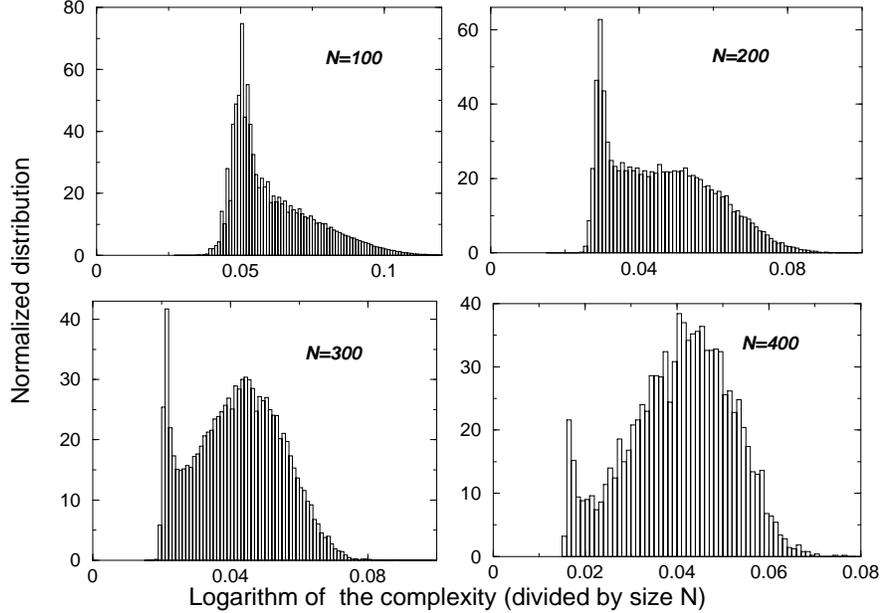,width=330pt,angle=-90}
\vskip .5cm
\caption{Probability distribution of the logarithm $\omega$ of the complexity  
(base 2, and divided by $N$) for $\alpha=3.5$ and for different sizes $N$.
Histograms are normalized to unity and obtained from 400,000 ($N=100$), 
50,000 ($N=200$), 20,000 ($N=300$), and  5,000 ($N=400$) samples}
\label{histolog}
\end{figure}
\end{center}

We have first performed some experiments to understand the
distribution of instance-to-instance fluctuations of the solving
times\cite{runtorun,runtorun2,distribution}. 
We draw randomly a large number of instances at fixed ratio 
$\alpha =3.5$ and size $N$ and, for each of them, run DPLL until
a solution is found (a small number of unsat instances can
be present and are discarded). We show in Figure~\ref{histolog}
the normalized histogram of the logarithms $\omega$
of the corresponding complexities $Q=2^{N\omega}$. 
The histogram is made of a narrow peak (left side) followed by a
wider bump (right side). As $N$ grows, the right peak acquires more
and more weight, while the left peak progressively disappears.
The abscissa of the center of the right peak gets slightly
shifted to the left, but seems to reach a finite value 
$\omega ^* \simeq 0.035$ as $N\to \infty$\cite{Coc1}. This right peaks
thus corresponds to the core of exponentially hard resolutions: w.h.p.
resolutions of instances require a time scaling as 
$2^{N \omega ^*}$ as the size of the instance gets larger and larger,
in agreement with the discussion of Section~II.

\begin{center}
\begin{figure}
\psfig{figure=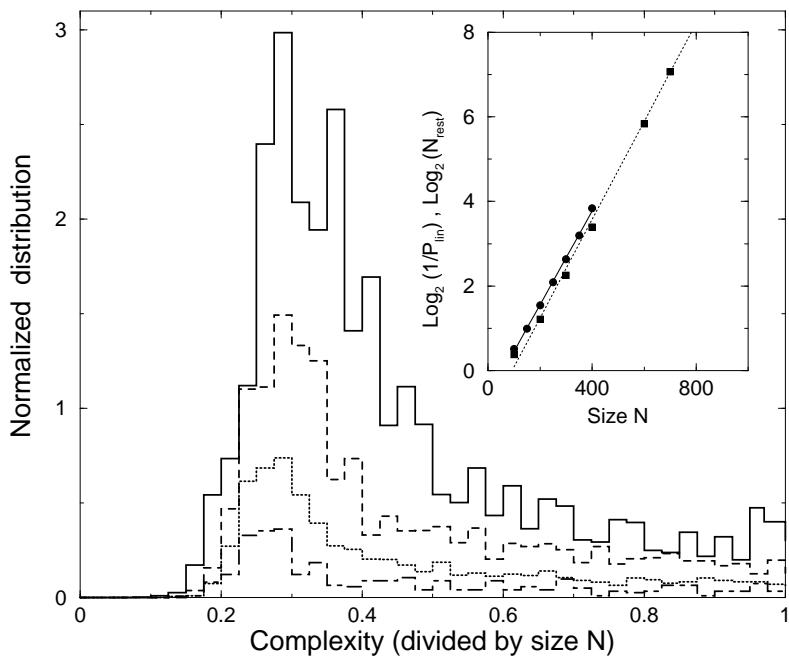,height=250pt,angle=-90}
\vskip .2cm
\caption{Probability distributions of the complexity $Q$ (divided by the
size $N$) for sizes $N=100$ (full line), $N=200$ (dashed line),
$N=300$ (dotted line), $N=400$ (dashed-dotted line). 
Distributions are not shown for complexities larger than $N$.  
Inset: Minus logarithm of the cumulative probability  
of complexities smaller or equal to $N$ as a function of $N$, for sizes
ranging from 100 to 400 (full line); logarithm of the number of
restarts necessary to find a solution for sizes ranging from 100 to
1000 (dotted line). Slopes are equal to $\zeta = 0.0011$ and $\bar
\zeta = 0.00115$ respectively.}
\label{histolin}
\end{figure}
\end{center}

On the contrary, the location of the maximum of the left peak
seems to vanish as $\log_2(N)/N$ when the size $N$ increases,
indicating that the left peak accounts for polynomial
(linear) resolutions. We have thus replotted the data shown in 
Figure~\ref{histolog}, changing the scale of the horizontal
axis $\omega = \log_2(Q)/N$ into $Q/N$. Results are shown in 
Figure~\ref{histolin}. We have limited ourself to $Q/N<1$, the range
of interest to analyse the left peak of Figure~\ref{histolog}. 
The maximum of the distribution is located at $Q/N\simeq 0.2-0.25$,
with weak dependence upon $N$. The cumulative probability $P_{lin}$ to have a 
complexity $Q$ less than, or equal to $N$, {\em i.e.} the integral of
Figure~\ref{histolin} over $0<Q/N<1$, decreases very quickly with $N$.
We find an exponential decrease, $P_{lin}= 2 ^{-N \zeta}$,
see Inset of Figure~\ref{histolin}. The rate $\zeta \simeq 0.011
\pm 0.001$ is determined  from the slope of the logarithm of the probability
shown in the Inset.  

\subsection{Locus of highest backtracking points.}

\begin{center}
\begin{figure}
\psfig{figure=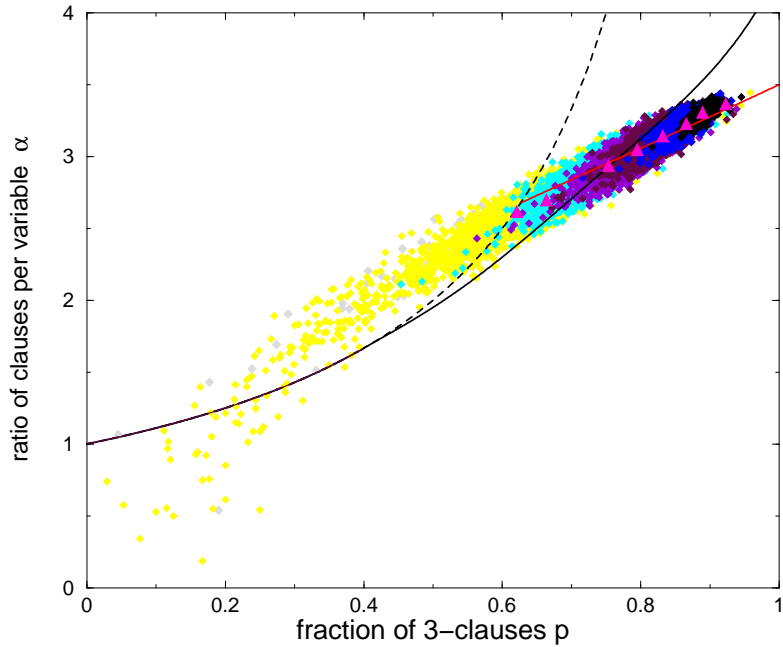,height=250pt,angle=-90}
\vskip .2cm
\caption{Locus of highest backtracking points $G$ in the phase diagram of
the 2+p-SAT model for 20,000 instances with $N=300$. 
The bold gray line represent the first branch trajectory for $\alpha=3.5$. 
Colors reflect the complexities of the instances, whose
logarithms $\omega$ range from 0.01 to 0.09, and are divided into 8
intervals of width $\Delta \omega=0.01$ and increasing darkness.
Filled triangles are the center of masses of points $G$ 
for each of the 8 intervals (the larger $\omega$,
the closer to the $p=1$ axis).}
\label{locus}
\end{figure}
\end{center}

To gain some intuition on the origin of fast, linear resolutions, we
have looked for the locus of the highest backtracking nodes $G$ in the
search trees. In the infinite size limit, $G$ is located w.h.p. at the
crossing $G^*$ of the resolution trajectory and the critical sat/unsat
line (Section~II). In Figure~\ref{locus} we show numerical evidence
for the link between complexity and trajectories in the phase diagram
for finite instance sizes.  We have run 20,000 instances ($\alpha=3.5,
N=300$), and reported for each of them the coordinates $p_G,\alpha_G$
of the highest backtracking point, and the logarithm $\omega$ of the
corresponding complexity. Large complexities ($\omega \ge 0.3$, right
bump of Figure~\ref{histolog}) are associated to points $G$ forming a
cloud centered around $G^*$ in the phase diagram of the 2+p-SAT model,
while points $G$ related to small complexities ($\omega \le 0.2$, left
peak of Figure~\ref{histolog}) are much more scattered in the phase
diagram. Notice the strong correlation between the value of $\omega$
and the average location of $G$ along the branch trajectory of Section
II. In the following we will concentrate on linear resolutions only. A
complementary analysis of the distribution of exponential 
resolutions for the problem of the vertex covering of random graphs 
was recently done by Montanari and Zecchina \shortcite{Mont}.

Figure~\ref{locus} shows that easy resolutions correspond to
trajectories capable of trespassing the contradiction line
$\alpha=1/(1-p)$. This, in addition to the linear scaling of the
corresponding complexities, indicates that easy resolutions coincide
with first descents in the search tree ending with a contradiction
located far beyond $D$ in the phase diagram, and then requiring a very
limited amount of backtracking before a solution is found.

This statement is supported by the analysis of the number of
unit-clause generated during easy resolutions. We have measured the
maximal number $(C_1)_{max}$ of unit-clauses generated along the last branch
in the tree, leading to the solution $S$ (Figure~\ref{tree}B).  
We found that $(C_1)_{max}$ scales linearly with $N$ with an extrapolated
ratio $(C_1)_{max}/N \simeq 0.022$ for $\alpha =3.5$. This linear scaling of
the number of unit-clauses is an additional proof of the trajectory 
entering the "dangerous" region $\alpha >1/(1-p)$ of the phase
diagram where unit-clauses accumulate. In presence of a $O(N)$ number
of 1-clauses, the probability of survival of the branch (absence of
contradictory literals among the unit-clauses) will be exponentially
small in $N$, in agreement with scaling of the left peak weight in
Figure~\ref{histolog}.

\subsection{Run-to-run fluctuations and restart experiments.}

\begin{center}
\begin{figure}
\psfig{figure=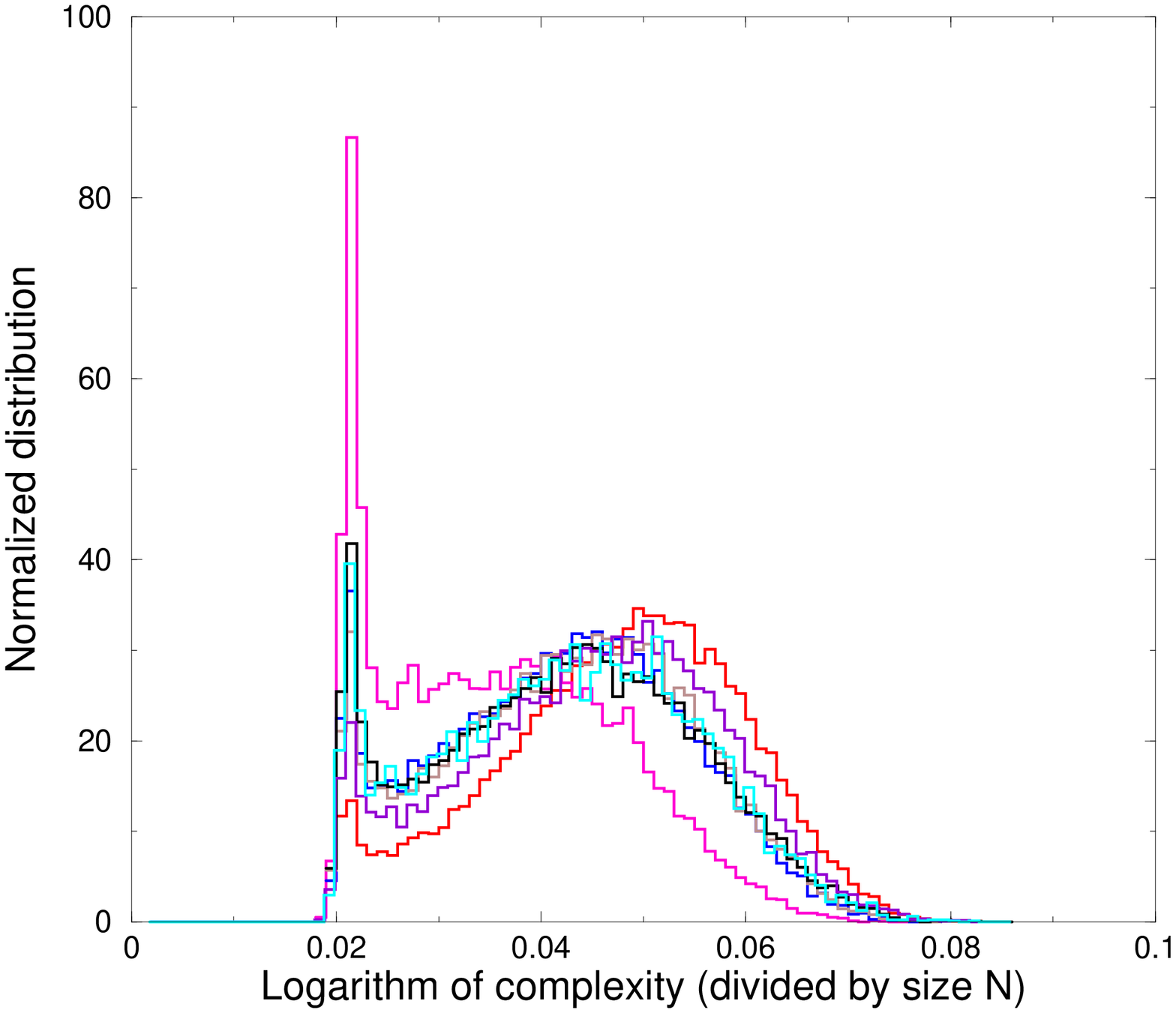,height=220pt,width=290pt,angle=-90}
\vskip .5cm
\caption{Probability distributions of the logarithm $\omega$ of the 
resolution complexity from 20,000 runs of DPLL. Each one of the five 
distribution corresponds 
to one randomly drawn instance of size $N=300$. The black curve is the 
instance-to-instance fluctuations of the complexity shown on Figure~3.}
\label{historun}
\end{figure}
\end{center}

We have so far considered the instance-to-instance fluctuations of the
complexity, that is the distribution of complexity obtained from
one run of DPLL on each of a large number of instances.
In Figure~\ref{historun}, we now show the histogram of complexities for
a large number of runs on a unique, random instance. After each run, 
clauses and variables are randomly relabeled to avoid any correlation
between different runs. Figure~\ref{historun} shows that these
run-to-run distributions are qualitatively independent of the particular
instances, and similar to the instance-to-instance distribution of
Figure~\ref{historun}\cite{runtorun,runtorun2}.  
%The location of the linear resolution peak is
%largely instance independent, while its weight seems to exhibit some
%dependence upon instance. 
    
The similarity between run-to-run and instance-to-instance
fluctuations for large sizes speaks up for the use of a systematic
stop-and-restart heuristic to speed up resolution: if a solution is
not found before $N$ splits, DPLL is stopped and launched again after
some random permutations of the variables and clauses.  Intuitively,
the expected number of restarts necessary to find a solution should
indeed be equal to the inverse of the weight of the linear complexity
peak in Figure~\ref{histolog}, with a resulting total complexity
scaling as $N \; 2^{\,0.011\,N}$, and much smaller than the one-run
complexity $2^{\,0.035 \, N}$ of DPLL (Section II).

We check the above reasoning by measuring the number $N_{rest}$ of
restarts performed before a solution is finally reached with the
stop-and-restart heuristic, and averaging $\log_2 (N_{res})$ over a
large number of random instances. Results are reports in the Inset of
Figure~\ref{histolin}.  The typical number $N_{rest}= 2^{N\bar \zeta}$
of required restarts clearly grows exponentially as a function of the
size $N$ with a rate $\bar \zeta = 0.0115 \pm 0.001$. To the accuracy of
the experiments, $\zeta$ and $\bar \zeta$ coincide as expected.

\section{Large deviation analysis of the first descent in the tree.}

\subsection{Evolution equation for the instance.}

Hereafter we compute the probability $\tilde P(C_1,C_2,C_3; T)$ that
the first branch of the tree carries an instance with $C_j$
$j$-clauses ($j=1,2,3$) after $T$ variables have
been assigned (and no contradiction has occurred). 
Let us call ${\bf C}$ the vector $(C_1,C_2,C_3)$.
$P$ obeys a Markovian evolution 
\begin{equation}
\tilde P({\bf C}; T+1) = \sum _{\bf C'}  K ({\bf C},{\bf C'}; T) 
\tilde P({\bf C'}; T)
\quad , \label{markov}
\end{equation}
where the entries of the transition matrix $K$ are equal to (for 
the GUC heuristic), 
\begin{eqnarray}
\label{bbra}
&& K ({\bf C},{\bf C'}; T) = {C_3' \choose C_3'-C_3} \; \left
(\frac 3{2M}\right)^{C_3'-C_3} \; \left(1-\frac 3M\right)^{C_3}
\times \nonumber \\
&& \sum_{w_2=0}^{C_3'-C_3} 
{C_3'-C_3\choose w_2}  \left \{(1 - \delta _{C'_1} )\;
\sum_{z_2=0}^{C_2'} {C_2' \choose z_2} \left (\frac
1{M}\right)^{z_2}\, \left (1- \frac 2M\right)^{C_2'-z_2} \right.
\times \nonumber \\ &&
\sum_{w_1=0}^{z_2} {z_2\choose w_1} \sum_{z_1=0}^{C_1'-1}  
{C_1'-1\choose z_1}\left (\frac {1} {2M}\right)^{z_1} 
\left( 1- \frac 1 M\right)^{C_1'-1-z_1}\; \times
\nonumber \\
&& \delta_{C_2-C_2'-w_2+z_2}
\;\delta_{C_1-C_1'-w_1+z_1+1}  +\; \delta_{C_1'} 
\sum_{z_2=0}^{C_2'-1} {C_2'-1 \choose z_2} \left (\frac
1{M}\right)^{z_2}\, 
\times \nonumber \\
&&\left( 1- \frac 2{M}\right)^{C_2'-1-z_2}
 \left. \sum_{w_1=0}^{z_2}  {z_2\choose w_1}\;
\delta_{C_2-C_2'-w_2+z_2+1} \; \delta_{C_1-w_1} 
 \right\} \ ,
\end{eqnarray}
and $\delta _C$ denotes the Kronecker delta function:
$\delta _C=1$ if $C=0$, $\delta _C=0$ otherwise. $M\equiv N-T$ is the
number of unassigned variables.

We then define the generating function $P(\,{\bf y}\,;T\,)$ of
the probabilities $\tilde P(\,{\bf C}\,;T\,)$ 
where ${\bf y}\equiv(y_1,y_2,y_3)$, through
\begin{equation}
\label{gener}
{P}(\,{\bf y}\,;T\,) =\sum_{\bf C}\; e^{\,{\bf y} \cdot 
{\bf C}}\; \tilde P(\,{\bf C}\,,T\,)\;
\end{equation}
where $\cdot $ denotes the scalar product. 
Evolution equation (\ref{markov}) can be rewritten in term of the 
generating function $P$,
\begin{eqnarray}
\label{eqev}
{P}(\,{\bf y}\,;T+1\,) &=& e^{-g_1({\bf y})} \; {P}\big(
\,{\bf g}({\bf y})\,; T\,\big) + \nonumber \\
&& \left ( e^{- g_2({\bf y})}- e^{- g_1({\bf y})}\right) \;
 {P}\big(-\infty,\,g_2({\bf y}),\,g_3({\bf y})\,; T\, \big)
\end{eqnarray}
where ${\bf g}$ is a vectorial function of argument ${\bf y}$ whose
components read
\begin{eqnarray}
g_1 ({\bf y}) &=&y_1+\ln\left[1+\frac 1 {N-T} \left
(\frac{e^{-y_1}}{2}-1\right)\right]
 \nonumber\\
g_2({\bf y})&=&y_2+\ln\left[1+\frac 2 {N-T} \left (\frac{e^{-y_2}}{2} 
\left(1+ e^{y_1}\right) -1\right)\right]
 \nonumber \\
g_3 ({\bf y})&=&y_3+\ln\left[1+\frac 3 {N-T}  \left (\frac{e^{-y_3}}{2} 
\left(1+ e^{y_2}\right) -1\right)\right]
\end{eqnarray}
We now solve equation (\ref{eqev}) by making some hypothesis on the scaling 
behavior of $P$ for large sizes.

\subsection{Hypothesis for the large $N$ scaling of the probability.}

Calculations leading to equation (\ref{eqev}) are rigorous. We shall
now make some hypothesis on $\tilde P, P$ that we believe to be correct
in the large size $N$ limit, but without providing rigorous proofs
for their validity. Our approach, common in statistical mechanics, may
be seen as a practical way to establish conjectures.

First, each time DPLL assigns variables through splitting or
unit-propagation, the numbers $C_j$ of clauses of length $j$ undergo
$O(1)$ changes. It is thus sensible to assume that after a number $T = t\,
N$ of variables have been assigned, the densities $c_j=C_j/N$ of
clauses have been modified by $O(1)$. This translates into a scaling
Ansatz for the probability $\tilde P$, 
\begin{equation}
\label{scalinghyp}
\tilde P({\bf C};T ) = e ^{ \,N \, \tilde \varphi ( c_1, c_2, c_3 ; t\,)} 
\qquad \qquad (\tilde \varphi \le 0 )
\end{equation}
up to non exponential in $N$ corrections. From equations (\ref{gener})
and (\ref{scalinghyp}), we obtain the following scaling hypothesis for
the generating function $P$,
\begin{equation}
\label{scalinghyp2}
P({\bf y};T ) = e ^{\, N \, \varphi ( \,{\bf y}\, ; t\,)} 
\end{equation}
up to non exponential in $N$ terms. Notice that $\varphi$ and $\tilde
\varphi$ are simply related to each other through Legendre transform,
\begin{eqnarray}
\varphi (\, {\bf y}\,;t\,) &=& \max _{\bf c} \bigg[ \tilde \varphi(\,
{\bf c} \, ; t\, ) + {\bf y} \cdot {\bf c}\bigg] \qquad , 
\\
\tilde \varphi (\, {\bf c}\,;t\,) &=& \min _{\bf y} \bigg[ \varphi(\,
{\bf y} \, ; t\, ) - {\bf y} \cdot {\bf c}\bigg] 
\qquad . \label{inversion}
\end{eqnarray}
In particular, $\varphi ( {\bf y}={\bf 0};t)$ is the logarithm of the
probability (divided by $N$) that the first branch has not been hit by
any contradiction after a fraction $t$ of variables have been
assigned. The most probable values of the densities $c_j(t)$ of 
$j$-clauses are then 
obtained from the partial derivatives of $\varphi(\,{\bf y}\,;t\,)$ in
${\bf y}={\bf 0}$:  $c_j(t)=\partial \varphi/\partial y_j ({\bf y}={\bf 0})$. 

We now present the partial differential equations (PDE) obeyed by $\varphi$.
Two cases must be distinguished: the number $C_1$ of
unit-clauses may be bounded ($C_1=O(1), c_1=o(1)$), or of 
the order of the instance size ($C_1=\Theta (N), c_1 = \Theta (1)$). 

\subsection{Case $C_1 = O(1)$: a large
deviation analysis around Chao and Franco's result.}

When DPLL starts running on a 3-SAT instance, very few unit-clauses
are generated and splittings occur frequently. In other words, the
probability that $C_1=0$ is strictly positive when $N$ gets
large. Consequently, both terms on the r.h.s. of (\ref{eqev}) 
are of the same order, and we make the hypothesis that $\varphi$ does
not depend on $y_1$: $\varphi (y_1,y_2,y_3;t) = \varphi (y_2,y_3;t)$.
This hypothesis simply expresses that 
$c_1=\partial \varphi/\partial y_1$ identically vanishes.

Inserting expression (\ref{scalinghyp2}) into the evolution equation 
(\ref{eqev}), we find\footnote{PDE (\ref{pde1}) is correct in the major
part of the $y_1,y_2,y_3$ space and, in particular, in the vicinity of
${\bf y}={\bf 0}$ we focus on in this paper. It has however to be 
to modified in a small region of the $y_1,y_2,y_3$ space; a complete
analysis of this case is not reported here but may be easily reconstructed
along the lines of  
Appendix A in  \cite{Coc2}.}
\begin{eqnarray} \label{pde1}
\frac{\partial \varphi}{\partial t} (y_2,y_3;t) = -y_2 &+& 2\, \gamma 
(y_2,y_2;t)\; \frac{\partial \varphi}{\partial y_2} (y_2,y_3;t)
\nonumber \\ &+& 3 \, \gamma (y_2,y_3;t)\;
\frac{\partial \varphi}{\partial y_3} (y_2,y_3;t)
\end{eqnarray}
where function $\gamma$ is defined through, 
\begin{equation}
\gamma (u,v;t) =\frac 1{1-t} \left (\frac{e^{-v}}{2} 
\left(1+ e^{u}\right) -1\right) \qquad .
\label{gammas}
\end{equation}
PDE (\ref{pde1}) together with initial condition $\varphi({\bf y};
t=0) = \alpha _0\, y_3$ (where $\alpha _0$ is the ratio of clauses per
variable of the 3-SAT instance) can be solved exactly with the resulting
expression,
\begin{eqnarray} \label{ld}
\varphi (y_2,y_3;t) &=& 
\alpha_0 \ln \left[ 1+ (1-t)^3 \left( e ^{y_3} - \frac 34 e ^{y_2} -
\frac 14 \right) + \frac {3 (1-t)}4 (e ^{y_2}-1) \right]
\nonumber \\  &+& 
(1-t)\, y_2\, e ^{y_2} + (1-t) (e ^{y_2}-1) \ln(1-t)
 \nonumber \\
&-&\left(e ^{y_2} +t-t \,e ^{y_2} \right) 
\ln \left( e ^{y_2} +t-t\, e ^{y_2} \right) 
\end{eqnarray}
Chao and Franco's analysis of the GUC heuristic may be recovered when
$y_2=y_3=0$ as expected. It is very easy to check that
$\varphi(y_2=0,y_3=0;t)=0$ (the probability of survival of the
branch is not exponentially small in $N$\cite{Fri}), and that the
derivatives $c_2(t), c_3(t)$ of $\varphi (y_2,y_3;t)$ with respect to
$y_2$ and $y_3$ coincide with the solutions of 
(\ref{diff}). In addition, our calculation provides also a complete
description of rare deviations of the resolution trajectory from its
highly probable locus shown on Figure~\ref{diag}. As a simple
numerical example, consider DPLL acting on a 3-SAT instance of ratio
$\alpha_0 = 3.5$. Chao and Franco's analysis shows that, once e.g.
$t=20\%$ of variables have been assigned, the densities of 2- and
3-clauses are w.h.p. equal to $c_2\simeq0.577$ and $c_3\simeq1.792$ 
respectively. Expression (\ref{ld}) gives access to the exponentially
small probabilities that $c_2$ and $c_3$ differ from their most
probable values. For instance, choosing $y_2=-0.1, y_3=0.05$, we find from
(\ref{ld}) and (\ref{inversion}) that there is a probability $e ^{-
0.00567 N}$ that $c_2 =0.504$ and $c_3=1.873$ for the same fraction $t=0.2$ of
eliminated variables.  By scanning all the values of $y_2,y_3$ we can
obtain a complete description of large deviations from Chao and 
Franco's result\footnote{Though we are not concerned here with 
subexponential (in $N$) corrections to 
probabilities, let us mention that it is also possible to calculate the
probability of split ($C_1=0$) per unit of time, extending Frieze and
Suen's result \shortcite{Fri} to ${\bf y}\ne {\bf 0}$.}.

The assumption $C_1= O(1)$ breaks down for the most probable
trajectory at some fraction $t_D$ e.g. $t_D\simeq 0.308$ for $\alpha
_0=3.5$ at which the trajectory hits point $D$ on Figure~\ref{diag}.
Beyond $D$, 1-clauses accumulate and the probability of survival of
the first branch is exponentially small in $N$.

\subsection{Case $C_1 = O(N)$: passing through the "dangerous" region.}

When the number of unit-clauses becomes of the order of $N$,
variables are a.s. assigned through unit-propagation. The first term
on the r.h.s. of equation (\ref{eqev}) is now exponentially dominant
with respect to the second one. The density of 1-clauses is strictly
positive, and $\varphi$ depends on $y_1$. We then obtain the following 
PDE,
\begin{eqnarray} \label{pde2}
\frac{\partial \varphi}{\partial t} (y_1,y_2,y_3;t) &=& -y_1 + \gamma  
(-\infty, y_1;t)\; \frac{\partial \varphi}{\partial y_1} (y_1,y_2,y_3;t)
\nonumber \\
&+& 2 \, \gamma   (y_1,y_2;t)\; \frac{\partial \varphi}{\partial y_2} 
(y_1,y_2,y_3;t) \nonumber \\
&+& 3 \, \gamma (y_2,y_3;t)\; \frac{\partial \varphi}
{\partial y_3} (y_1,y_2,y_3;t)
\end{eqnarray}
with $\gamma (u,v;t)$ given by equation (\ref{gammas}).
When $y_1=y_2=y_3=0$, equation (\ref{pde2}) simplifies to
\begin{equation} \label{pde3}
\frac{d z}{d t} (t) = -\frac{c_1(t)}{2(1-t)}
\qquad,
\end{equation}
where $c_1(t)$ is the most probable value of the density of unit-clauses,
and $z(t)$ is the logarithm of the probability that the branch has
not encountered any contradiction (divided by $N$).
The interpretation of (\ref{pde3}) is transparent. Each time a literal
is assigned through unit-propagation, there is  a probability
$(1-1/2(N-T))^{C_1-1} \simeq e ^{-c_1/2/(1-t)}$ that no contradiction occurs.
The r.h.s. of (\ref{pde3}) thus corresponds to the rate of decay of
$z$  with "time" $t$.

We have not been able to solve analytically PDE (\ref{pde2}), and have
resorted to an expansion of $\varphi$ in powers of ${\bf y}$.
To $k^{th}$ order, we approximate the solution of
(\ref{pde2}) by a polynomial of total degree $k$,
\begin{equation} \label{app}
\varphi ^{(k)} ({\bf y};t) = \sum _{e_1+e_2+e_3 \le k} 
\varphi ^{(k)} _{e_1,e_2,e_3} (t) \; y_1 ^{e_1} \; y_2 ^{e_2} \; y_3 ^{e_3} 
\end{equation}
and insert (\ref{app}) on the r.h.s. of (\ref{pde2}). We collect on
the l.h.s. the terms of degrees $\le k$ and obtain a set of ${\cal
N}_k=(k+3)(k+2)(k+1)/6$ first order coupled
linear ODEs for the coefficients $\varphi ^{(k)} _{e_1,e_2,e_3} (t)$ of the 
polynomial (\ref{app}). This approximation gets better and better as
$k$ increases at a cost of more and more coupled ODEs to be solved.
The initial conditions for these ODEs are chosen to match the expansion of
the exact solution (\ref{ld}) at time $t_D$. 

\begin{table}
\begin{tabular}{ccccccc} 
\hline
{\hbox{\rm order}}&  
{\hbox{\rm $\zeta$}} & {\hbox{$(c_1) _{max}$}}  &{\hbox {\rm$
t_{D'}$ }} &
 {\hbox{$p_{D'}$}}& {\hbox {$\alpha_{D'}$}}&{\hbox {{$\gamma$}}}\\
\hline  
1   &      .0384 &  .0502  & .8878 &  .0804 &   .5477 &   .1720 \\
2   &      .0036 &  .0121  & .6553 &  .2707 &    1.575 &  .1990  \\
3   &      .0098 &  .0227  & .7495 &  .1901 &    1.201&   .2069  \\
4   &      .0098 &  .0226  &  .7483 & .1911 &    1.206&   .2069 \\
\hline
\end{tabular}
\caption{Results at different orders $k$ of approximation
for $\alpha _0=3.5$: 
logarithm $\zeta$ of the probability that the first branch is not hit by any 
contradiction, maximal density $(c_1)_{max}$ of unit-clauses ever
reached, fraction of eliminated variables $t_{D'}$
and coordinates $p_{D'}, \alpha_{D'}$  at point $D'$ {\em i.e.} when
the number of unit-clauses ceases to be $O(N)$,
complexity ratio $\gamma = Q/N$ of the corresponding linear
resolution.} \label{table}
\end{table}

At the lowest order ($k=1$), we find a set of four coupled equations
for $z^{(1)}(t) \equiv \varphi ^{(1)} _{0,0,0} (t), c_1^{(1)}(t)
\equiv \varphi ^{(1)} _{1,0,0} (t), c_2^{(1)}(t)\equiv \varphi
^{(1)} _{0,1,0} (t), c_3^{(1)}(t)\equiv \varphi
^{(1)} _{0,0,1} (t)$ that read
\begin{eqnarray}
\frac {d c_1 ^{(1)}(t)} {dt} &=& -\frac{c_1^{(1)} (t)}{2(1-t)} 
+  \frac{c_2 ^{(1)}(t)}{1-t} \nonumber \\
\frac {d c_2 ^{(1)}(t)} {dt} &=&  -\frac{2\,c_2 ^{(1)}(t)}{1-t}
+\frac{3\,c_3 ^{(1)}(t)}{2(1-t)}  \nonumber \\
\frac {d c_3 ^{(1)}(t)} {dt} &=& - \frac{3 \, c_3 ^{(1)}(t)}{1-t}
\nonumber \\
\frac {d z ^{(1)}(t)} {dt} &=&  - \frac{c_1 ^{(1)}(t)}{2(1-t)} 
\quad , \label{diff2}
\end{eqnarray}
together with the initial conditions $c_1  ^{(1)}(t_D) = z
^{(1)}(t_D)=0, c_2  ^{(1)}(t_D) = 1-t_D, c_2  ^{(1)}(t_D) = \alpha _0
(1-t_D)^3$, with $t_D$ uniquely determined from $\alpha_0$. 
The solution of (\ref{diff2}) for $\alpha _0 =3.5$
shows that $c_1$ first increases and
reaches its top value $(c_1^{(1)})_{max}\simeq 0.05$. It then decreases
and vanishes at $t^{(1)}_{D'}\simeq 0.89$, where the trajectory exits
the "dangerous" region  where contradictions occurs
w.h.p. (Figure~\ref{diag}). The probability of this event scales as 
$2^{- N \zeta ^{(1)} }$ for large $N$, with $\zeta ^{(1)} = -
z ^{(1)}(t^{(1)}_{D'})/\ln 2 \simeq 0.038$. The end of the resolution
trajectory obeys Chao and Franco's equations (\ref{diff}).

Results improve when going to higher orders in $k$, see
Table~\ref{table}. No sensible difference can be found between $k=3$ and
$k=4$ results. The calculated values of $\zeta \simeq 0.01, (c_1)
_{max}\simeq 0.022$ and $\gamma \simeq 0.21$ are in very good
agreement with the numerical experiments of Section~III.

\begin{center}
\begin{figure}
\psfig{figure=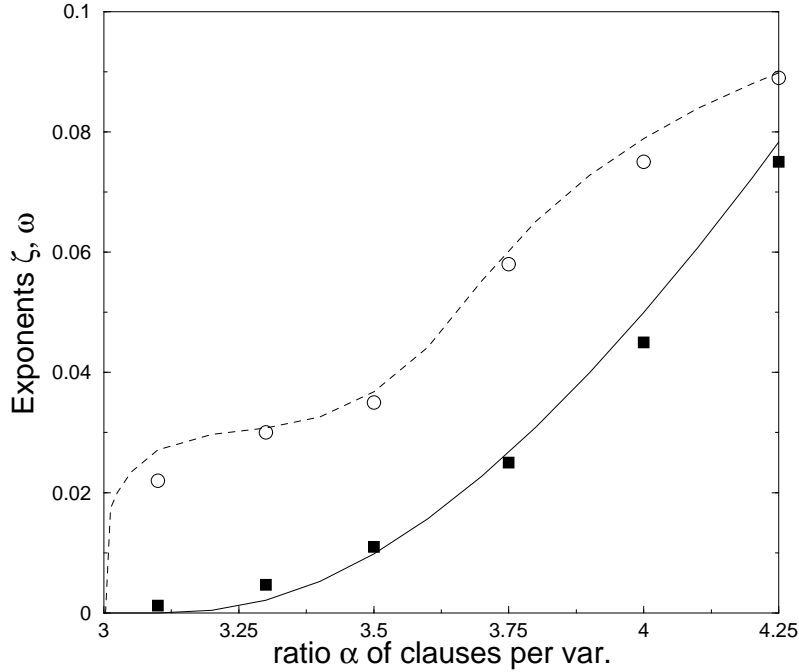,height=260pt,angle=-90}
\vskip .5cm
\caption{Exponent $\zeta$ of the linear resolution probability
(simulations: filled squares, theory: full
line), and exponent $\omega$ of the typical complexity
(simulations: empty circles, theory from (Cocco, Monasson 2001b): 
dotted line), as a function of the clause per variable ratio $\alpha$.}
\label{autre}
\end{figure}
\end{center}

We report on Figure~\ref{autre} the experimental and theoretical
values of $\zeta$ found over the whole range $\alpha _L \le \alpha _0
\le \alpha_C$. Note the very good agreement between our
quantitative theory and simulations, which supports the scaling 
hypothesis made above. 

\section{Conclusions.}

In this work, we have studied deviations from the typical ({\em i.e.}
occurring w.h.p.)  solving complexity of satisfiable random 3-SAT
instances using DPLL algorithm with a simple splitting heuristic (GUC)
\cite{fra2,Fra}. For ratios $\alpha$ of clauses per variable in the
range $\alpha_L=3.003<\alpha<\alpha_C$, complexity grows almost surely
exponentially with the size $N$ of the instance, but resolution may
very rarely ({\em i.e.}  with an exponentially small probability)
require a polynomial (linear) computational effort only.  These linear
resolutions correspond to search tree reducing to a single branch
essentially (Figure~\ref{tree}A), and can be visualized as
trajectories that cross the unsat phase of the Figure~\ref{diag}
diagram without being stopped by any contradiction.  Our approach
allowed us to calculate the large deviations from typical
resolutions, and the exponent $\zeta$ of the probability
$\tilde P \sim 2^{-\zeta\, N}$ of linear resolutions. Our
theoretical calculation predicts for instance that the exponent
corresponding to random 3-SAT instances with ratio $\alpha =3.5$
equals $\zeta \simeq 0.01$, in very good agreement with the values
extrapolated from the histogram of resolution time on different
instances (instance-to-instance distribution of complexities) and the
value $\bar \zeta$ extrapolated from the number of restarts necessary
to solve one random instance (Inset of Figure~\ref{histolin}).

The computational effort to find a solution with the systematic
restart procedure, $N_{rest} \sim 1/ P _{lin} \sim 2^{N \zeta}$ turns out
to be exponentially smaller than the typical time to find a solution 
$2^{N \omega}$ without restart (e.g. $\omega=0.035$ for $\alpha
=3.5$). Our calculation gives thus some theoretical support to the use
of restart-like procedures (see also \cite{Mont} for recent
theoretical results), empirically known to speed up considerably 
resolutions\cite{Dubois,Gomes}. To be more concrete, while, without restarts,
we were able to solve with DPLL algorithm instances with 500
variables in about one day of CPU (for $\alpha =3.5$), the restart
procedure allows to solve instances with 1000 variables in 15 minutes 
with the same computer and splitting heuristic (GUC).

The present work suggests that the cut-off time, at
which the search is halted and restarted, need not be precisely tuned
but is simply given by the size of the instance. This conclusion could
be generic and apply to other combinatorial decision
problems and other heuristics. More precisely, if a combinatorial
problem admits some efficient (polynomial) search heuristic for some
values of control parameter (e.g. the ratio $\alpha$ here, or the
average adjacency degree for the coloring problem of random graphs),
there might be an exponentially small probability that the heuristic
is still successful (in polynomial time) in the range of parameters
where resolution almost surely requires massive backtracking and exponential
effort. When the decay rate of the polynomial time 
resolution probability $\zeta$ is smaller than the growth rate
$\omega$ of the typical exponential resolution time, stop-and-restart 
procedures with a cut-off in the search equal to a polynomial of the 
instance size will lead to an exponential speed up of resolutions.

It would be interesting to extend the previous approach to more
sophisticated and powerful search e.g. satz of chaff heuristics.
It is however not clear how a full analytical study could be worked out
without resorting to approximate expressions for the transition matrix.
Another natural extension of the present work would be to focus on
other decisions problems e.g. graph coloring for which 
the high probability behaviour of simple heuristics is well understood
\cite{Achl4}. 

\acknowledgements
R. M. is supported in part by the ACI Jeunes Chercheurs 
``Algorithmes d'optimisation et syst{\`e}mes d{\'e}sordonn{\'e}s quantiques'' 
from the French Ministry of Research.

\end{article}

\begin{thebibliography}{}
\bibitem[\protect\citeauthoryear{Achlioptas and Molloy}{1997}]{Achl4}
Achlioptas, D. and Molloy, M.
Analysis of a List-Coloring Algorithm on a Random Graph,
in {\em Proceedings of FOCS 97}, p.204-212 (1997).

\bibitem[\protect\citeauthoryear{Achlioptas et al.}{2001a}]{Achl1}
Achlioptas, D., Kirousis, L., Kranakis, E. and Krizanc, D. 
Rigorous results for random (2+p)-SAT, {\em Theoretical Computer
Science} {\bf 265}, 109-129  (2001).

\bibitem[\protect\citeauthoryear{Achlioptas}{2001b}]{Achl}
Achlioptas, D.
Lower bounds for random 3-SAT via differential equations, 
{\em Theoretical Computer Science} {\bf 265}, 159--185 (2001).

\bibitem[\protect\citeauthoryear{Achlioptas, Beame and Molloy}{2001c}]{Achl3}
Achlioptas, D., Beame, P. and Molloy, M. 
A Sharp Threshold in Proof Complexity. 
in {\em Proceedings of STOC 01}, p.337-346 (2001).

\bibitem[\protect\citeauthoryear{Beame et al.}{1998}]{Bea}
Beame, P., Karp, R., Pitassi, T. and Saks, M.       
{\em ACM Symp.\ on Theory of Computing (STOC98)}, 561--571
Assoc. Comput. Mach., New York (1998).

\bibitem[\protect\citeauthoryear{Chao and Franco}{1986}]{fra2}
Chao, M.T. and Franco, J. 
Probabilistic analysis of two heuristics for the 3-satisfiability problem, 
{\em SIAM Journal on Computing} {\bf 15}, 1106-1118 (1986).

\bibitem[\protect\citeauthoryear{Chao and Franco}{1990}]{Fra}
Chao, M.T. and Franco, J.
Probabilistic analysis of a generalization of the unit-clause literal
selection heuristics for the k-satisfiability problem,
{\em Information Science} {\bf 51}, 289--314 (1990).

\bibitem[\protect\citeauthoryear{Chv{\`a}tal and Szmeredi}{1988}]{Chv}
Chv{\`a}tal, V. and Szmeredi, E. Many hard examples for resolution,    
{\em Journal of the ACM} {\bf 35}, 759--768 (1988).

\bibitem[\protect\citeauthoryear{Coarfa et al.}{2000}]{Vardi1}
Coarfa, C., Dernopoulos, D.D., San Miguel Aguirre, A., Subramanian, D.
and Vardi, M.Y. Random 3-SAT: The plot thickens. In R. Dechter, editor,
{\em Proc. Principles and Practice of Constraint Programming
(CP'2000)}, Lecture Notes in Computer Science 1894, 143-159 (2000);

\bibitem[\protect\citeauthoryear{Cocco and Monasson}{2001}]{Coc1}
Cocco, S. and Monasson R.
Trajectories in phase diagrams, growth processes 
and computational complexity: how search algorithms solve the 
3-Satisfiability problem, {\em Phys. Rev. Lett.} {\bf 86}, 1654 (2001)

\bibitem[\protect\citeauthoryear{Cocco and Monasson}{2001b}]{Coc2}
Cocco, S. and Monasson, R. 
Analysis of the computational complexity of solving random satisfiability 
problems using branch and bound search algorithms, {\em
Eur. Phys. J. B} {\bf 22}, 505 (2001).

\bibitem[\protect\citeauthoryear{Crawford and Auton}{1996}]{Cra}
Crawford, J. and Auton, L. 
Experimental Results on the Cross-Over Point in Satisfiability Problems,
{\em Proc.\ 11th Natl.\ Conference on Artificial Intelligence 
(AAAI-93),} 21--27, The AAAI Press / MIT Press, Cambridge, MA (1993);
{\em Artificial Intelligence} {\bf  81} (1996).

\bibitem[\protect\citeauthoryear{Davis, Logemann and Loveland}{1962}]{DP}
Davis, M., Logemann, G. and Loveland, D. 
{A machine program for theorem proving.}  
{\em Communications of the ACM} {\bf 5}, 394-397 (1962).

\bibitem[\protect\citeauthoryear{Dubois et al.}{1993}]{Dubois}
Dubois, O., Andre, P., Boufkhad, Y. and Carlier, J.
SAT versus UNSAT, {\em DIMACS Series in Discrete Math. and Computer 
Science},  415--436 (1993).

\bibitem[\protect\citeauthoryear{Dubois et al.}{2001}]{Dubtcs}
Dubois, O., Monasson, R., Selman, B. and Zecchina, R. (eds)
Phase transitions in combinatorial problems.
{\em Theor. Comp. Sci.} {\bf 265} (2001).

\bibitem[\protect\citeauthoryear{Franco}{2001}]{Francotcs}
Franco, J.
Results related to thresholds phenomena research in satisfiability:
lower bounds. 
{\em Theoretical Computer Science} {\bf 265}, 147--157 (2001).

\bibitem[\protect\citeauthoryear{Friedgut}{1999}]{Friedgut}
Friedgut, E.
Sharp thresholds of graph properties, and the k-sat problem,
{\em Journal of the A.M.S.} {\bf 12}, 1017 (1999).
 
\bibitem[\protect\citeauthoryear{Frieze and Suen}{1996}]{Fri}
Frieze, A. and Suen, S.
Analysis of two simple heuristics on a random instance of k-SAT,      
{\em Journal of Algorithms} {\bf 20}, 312--335 (1996).

\bibitem[\protect\citeauthoryear{Gent and Walsh}{1994}]{distribution}
Gent, I.P. and Walsh, T. Easy problems are sometimes hard,
{\em Artificial Intelligence} {\bf 70}, 335-345 (1994).

\bibitem[\protect\citeauthoryear{Gent, van Maaren and Walsh}{2000}]{Hans}
Gent, I., van Maaren, H. and Walsh, T. (eds),
SAT2000: Highlights of Satisfiability Research in the Year 2000, 
{\em Frontiers in Artificial Intelligence and Applications},
vol. 63, IOS Press, Amsterdam (2000).

\bibitem[\protect\citeauthoryear{Gomes et al.}{2000}]{Gomes}
Gomes, C.P., Selman, B., Crato, N. and Kautz, H. {\em J. Automated
Reasoning} {\bf 24}, 67 (2000).

\bibitem[\protect\citeauthoryear{Gu, Purdom, Franco and Wah}{1997}]{survey}
Gu, J., Purdom, P.W.,  Franco, J. and Wah, B.W. 
Algorithms for satisfiability (SAT) problem: a survey.
{\em DIMACS Series on Discrete Mathematics 
and Theoretical Computer Science} {\bf 35}, 19-151, 
American Mathematical Society (1997). 
 
\bibitem[\protect\citeauthoryear{Hartmann and Weigt}{2001}]{Wei}
Hartmann, A. and Weigt, M. 
Typical solution time for a vertex-covering algorithm on 
finite-connectivity random graphs, 
{\em Phys. Rev. Lett.} {\bf 86}, 1658 (2001).

\bibitem[\protect\citeauthoryear{Hogg and Williams}{1994}]{runtorun}
Hogg, T. and Williams, C.P. The hardest constraint problems: a double
phase transition, {\em Artificial Intelligence} {\bf 69}, 359-377
(1994).

\bibitem[\protect\citeauthoryear{Hogg, Huberman and Williams}{1996}]{AI}
Hogg, T., Huberman, B.A. and Williams, C. (eds)
Frontiers in problem solving: phase transitions and complexity.
{\em Artificial Intelligence} {\bf 81} I \& II (1996).

\bibitem[\protect\citeauthoryear{Kirkpatrick and Selman}{1994}]{Kir}
Kirkpatrick, S., and Selman, B.  
Critical Behavior in the Satisfiability of Random Boolean Expressions.
{\em Science} {\bf 264}, 1297--1301 (1994).

\bibitem[\protect\citeauthoryear{Mitchell, Selman and Levesque}{1992}]{Mit}
Mitchell, D., Selman, B. and Levesque, H.
Hard and Easy Distributions of SAT Problems,
{\em Proc.\ of the Tenth Natl.\ Conf.\ on Artificial Intelligence
(AAAI-92)}, 440-446,
The AAAI Press / MIT Press, Cambridge, MA (1992).

\bibitem[\protect\citeauthoryear{Monasson et al.}{1999}]{Sta}
Monasson, R., Zecchina, R., Kirkpatrick, S., Selman, B. and Troyansky, L.
Determining computational complexity from characteristic 'phase transitions'.
{\em Nature} {\bf 400}, 133--137 (1999);
2+p-SAT: Relation of Typical-Case Complexity to the Nature of
the Phase Transition,
{\em Random Structure and Algorithms} {\bf 15}, 414 (1999).

\bibitem[\protect\citeauthoryear{Montanari and Zecchina}{2002}]{Mont}
Montanari, A., and Zecchina, R. Boosting search by rare events,
{\em Phys. Rev. Lett.} {\bf 88}, 178701 (2002). 

\bibitem[\protect\citeauthoryear{San Miguel Aguirre et al.}{2001}]{Vardi2}
San Miguel Aguirre, A. and Vardi, M.Y. Random 3-SAT and BDDs: The plot
thickens further, (CP'2001) (2001).

\bibitem[\protect\citeauthoryear{Selman and Kirkpatrick}{1996}]{runtorun2}
Selman, B. and Kirkpatrick, S.
Critical behavior in the computational cost
of satisfiability testing, {\em Artificial Intelligence} {\bf 81},
273-295 (1996).
\end{thebibliography}
\end{document}